\documentclass[conference]{IEEEtran}
\IEEEoverridecommandlockouts

\usepackage{cite}
\usepackage{amsmath,amssymb,amsfonts}
\usepackage{algorithmic}
\usepackage{graphicx}
\usepackage{textcomp}
\usepackage{xcolor}
\def\BibTeX{{\rm B\kern-.05em{\sc i\kern-.025em b}\kern-.08em
    T\kern-.1667em\lower.7ex\hbox{E}\kern-.125emX}}

\usepackage{graphicx}

\usepackage{tikz}
\usepackage{amsmath}
\usepackage{comment}
\usepackage{booktabs}
\usepackage{url}
\usepackage{multicol}
\usepackage[breakable]{tcolorbox}

\usepackage{graphicx}
\usepackage{algorithmic}
\usepackage{subcaption}
\usepackage{todonotes}
\usepackage{hyperref}

\usepackage{graphicx}
\usepackage{xspace}

\definecolor{amber}{rgb}{1.0, 0.75, 0.0}

\usepackage{xcolor}

\newcommand{\Scam}{\texttt{Scam}\xspace}
\newcommand{\Clean}{\texttt{Clean}\xspace}
\newcommand{\Uncertain}{\texttt{Uncertain}\xspace}

\begin{document}

\title{Measuring and Evaluating the Performance of Generative AI Models for Scam Detection}

\author{
\IEEEauthorblockN{Cem Topcuoglu}
\IEEEauthorblockA{
\textit{Northeastern University}\\
Boston, MA, USA \\
topcuoglu.c@northeastern.edu}
\and
\hspace{-0.5cm}\IEEEauthorblockN{Seyed Ali Akhavani}
\IEEEauthorblockA{
\hspace{-0.5cm}\textit{Northeastern University}\\
\hspace{-0.5cm}Boston, MA, USA \\
\hspace{-0.5cm}sadatakhavani.s@northeastern.edu}
\and
\hspace{-0.1cm}\IEEEauthorblockN{Harel Berger}
\IEEEauthorblockA{
\hspace{-0.1cm}\textit{Ariel University}\\
\hspace{-0.1cm}Israel\\
\hspace{-0.1cm}harelb@ariel.ac.il}
\and
\IEEEauthorblockN{Sadia Afroz}
\IEEEauthorblockA{
\textit{Gen Digital}\\
USA \\
sadia.afroz@gendigital.com}
\and
\hspace{0.5cm}\IEEEauthorblockN{Michalis Pachilakis}
\IEEEauthorblockA{
\hspace{0.5cm}\textit{Gen Digital}\\
\hspace{0.5cm}Ireland \\
\hspace{0.5cm}michalis.pachilakis@gendigital.com}
\and
\IEEEauthorblockN{Vibhor Sehgal}
\IEEEauthorblockA{
\textit{Gen Digital}\\
USA \\
vibhor.sehgal@gendigital.com}
\and
\IEEEauthorblockN{Leyla Bilge}
\IEEEauthorblockA{
\textit{Gen Digital}\\
France \\
leylya.yumer@gendigital.com}
\and
\IEEEauthorblockN{Engin Kirda}
\IEEEauthorblockA{
\textit{Northeastern University}\\
Boston, MA, USA \\
ek@ccs.neu.edu}
}

\maketitle

\begin{abstract}
Online scams continue to cause substantial financial and personal harm. As a
result, detection systems based on Large Language Models (LLMs) have been
integrated into security products ranging from email gateways and
browser extensions to fraud-monitoring dashboards. As this adoption accelerates,
a common belief has taken hold: that these models are broadly suitable for scam
detection. In this work, we investigate whether LLMs, with their strong
capabilities in understanding intent, context, and reasoning, can effectively
detect scams across diverse scenarios without task-specific fine-tuning. We
curate and release a unique benchmark dataset of real-world scams
spanning multiple formats and topics. We evaluate nine LLMs of varying sizes and
architectures, examining their performance under different prompting strategies
and comparing them to a fine-tuned BERT-based classifier. Our results show that
while larger LLMs generally outperform smaller ones, effective prompting
substantially boosts the performance of smaller models. Moreover, LLMs are
better at generalizing to unseen scams compared to fine-tuned models, suggesting
that pre-trained knowledge contributes meaningfully to scam detection. We
release our dataset and evaluation framework to facilitate future research in
robust scam detection using language models.
\end{abstract}


\section{Introduction}

Online scams, unfortunately, are a growing threat. Scammers often exploit human
trust and emotions by mimicking trusted entities or triggering fear and greed.
Pig-butchering scams, for instance, start with a misdirected message and
escalate to fake investments, costing over \$75M
(2020-2024)~\cite{pig_butchering,griffin2024crypto}. Losses remain severe: in
2024 alone, the ACCC reported \$318M AUD~\cite{scamwatch_stats} and the FTC
reported \$12B USD~\cite{ftc}. Consequently, industry and academia are racing to
detect scams and reduce harm.

Traditional machine learning has been widely applied in scam detection. Some of
these techniques are Logistic Regression~\cite{sharifi2011detection},
SVM~\cite{Wu2019WhoAT}, Naive Bayes~\cite{C2024ScamCD}, and Suffix
Trees~\cite{chen2014scam}. However, these models, trained on known attacks,
generalize poorly to novel scams. As a result, detection systems based on Large
Language Models (LLMs) have been showing up in more and more security products
ranging from email gateways and browser extensions to fraud-monitoring
dashboards. As this adoption accelerates, a common belief has taken hold: that
these models are broadly suitable for scam detection. However, generative AI is
promising yet under-evaluated for this task, despite successes in domains such
as vulnerability~\cite{299549}, bug~\cite{299699}, and harmful content
detection~\cite{li2024hot}. Early efforts include an interactive scam-analysis
chatbot~\cite{genie} and KnowPhish's LLM-based website analysis~\cite{299740}.
Despite its recent popularity, a rigorous assessment of LLMs for the purpose of
scam detection is missing.

In this paper, we provide the first comprehensive empirical study of LLMs for
scam detection. Using 2,742 unique samples from three sources, we evaluate
Mistral (7B, 8X7B), Llama 3 (8B), Llama 3.1 (8B, 70B), Llama 3.2 (11B), and
OpenAI ChatGPT (4, 4o, 4o1) under zero/few-shot and chain-of-thought prompting.
Larger models perform better: ChatGPT 4o1 and Llama 3.1 70B achieve 64\% and
65\% micro-F1/precision/recall, respectively. Prompting matters but is
model-dependent: Llama 3.1 8B rises from 50\% to 57\% with tailored prompts,
while ChatGPT 4o1 can degrade under the same strategy. More shots and small
temperature/top-p adjustments yield modest gains. Category-level
precision/recall vary by model, informing deployment trade-offs. A fine-tuned
BERT trails the top LLMs and struggles on unseen samples, whereas LLMs
generalize better across datasets.

\noindent\textbf{Summary.} We make the following contributions:
\begin{itemize}
    \item We present the first study that evaluates the effectiveness and limitations of popular LLMs for scam detection by using various prompting techniques and hyperparameters.
    \item We fine-tune Mistral 7B, Llama 3.1 8B, and BERT models and evaluate their performance in scam detection.
    \item We empirically show that current LLMs have limitations and are not a silver-bullet solution for scam detection.
    \item We release the first benchmark for evaluating LLMs for scam detection, the ScamBenchmark dataset.

\end{itemize}

\textbf{Availability.} We open source datasets and fine-tuned models~\footnote{\url{https://github.com/cemtopcuoglu/genai_scam_detection}}.

\section{Background and Related Work}
\label{sec:background}

\subsection{Large Language Models}

Large Language Models (LLMs) are pre-trained on diverse datasets to learn
linguistic patterns, context, and semantics. Built on transformer
architecture~\cite{vaswani2017attention}, they process text in parallel,
enabling scalability and efficiency. LLMs are typically distinguished by
parameter count and training data scope. Larger models capture more nuanced
information but demand more resources; smaller models require precise prompting
to approach comparable performance.

Key factors influencing LLM behavior include training data quality, temporal
coverage, multilinguality, and internet access. Inference behavior is further
shaped by hyperparameters: temperature controls randomness (lower values yield
more deterministic outputs), and top-p limits token sampling to high-probability
candidates. Together, these determine adaptability and task suitability.

\noindent \textbf{BERT.} BERT (Bidirectional Encoder Representations from
Transformers)~\cite{devlin2018bert}, introduced by Google in 2018,
revolutionized NLP via deep bidirectional pre-training. BERT-base (110M) and
BERT-large (340M) were trained using masked language modeling and next-sentence
prediction, setting new benchmarks in transfer learning. Its success spurred
optimized variants such as RoBERTa, ALBERT, and DistilBERT.

\noindent \textbf{ChatGPT.} Launched by OpenAI in late 2022, ChatGPT extended
the GPT family to conversational contexts, excelling at multi-turn dialogue.
While proprietary, it is built on models such as GPT-3
(175B)~\cite{brown2020language}, with later iterations presumed larger. Its
release marked a major leap in accessible dialogue systems.

\noindent \textbf{LLaMA.} Meta's LLaMA~\cite{touvron2023llama}, released in 2023, democratized high-performance LLMs via open access. Offered in
7B–65B sizes, even the 13B model outperformed GPT-3 on many tasks. The 65B
version rivaled state-of-the-art systems, while the lightweight 7B/13B models
gained popularity for their performance–efficiency balance and modifiability.

\noindent \textbf{Mistral.} Mistral AI’s 2023 model
suite~\cite{jiang2023mistral} prioritized efficiency through architectural
innovation. Mistral-7B introduced sliding window and grouped-query attention,
outperforming larger models like LLaMA-13B. The Mixtral-8x7B mixture-of-experts
model extended this efficiency at scale, accelerating adoption across research
and industry.

\subsection{Prompt Engineering}

Prompts are structured inputs that steer LLMs toward specific outputs. Effective
prompt design is critical for controlling model behavior. Prompting techniques
vary by complexity and whether examples are provided. Below are key types:

\noindent \textbf{Zero-Shot Prompting.}
In zero-shot prompting, the model receives only an instruction, relying entirely on pre-trained knowledge to complete the task~\cite{brown2020language}.
\vspace{-0.4em}
\begin{quote}
\textbf{Example:} \\
Classify the following input as ``Scam,'' ``Clean,'' or ``Uncertain.'' \\
\textbf{Input:} \textit{INPUT}
\end{quote}
\vspace{-0.4em}
Best for straightforward tasks, zero-shot performance may decline on tasks requiring nuanced or contextual understanding.

\noindent \textbf{Few-Shot Prompting (K-Shot).}
Few-shot prompting provides examples that demonstrate the task. This helps the model identify patterns and improves accuracy, especially on ambiguous inputs~\cite{brown2020language,lu2021fantastically}.
\vspace{-0.4em}
\begin{quote}
\textbf{Example:} \\
- \textit{Confirm your bank details to keep your account active.} → Scam \\
- \textit{Your statement is ready. Log into your account.} → Clean \\
\textbf{Input:} \textit{INPUT}
\end{quote}
\vspace{-0.4em}
The number of examples (`K`) directly affects performance; even a single example can help.

\noindent \textbf{Chain-of-Thought (CoT) Prompting.}
CoT prompting guides the model to reason step-by-step before reaching a conclusion. This method improves performance on tasks involving logic, intent, or implicit cues~\cite{wei2022chain}.
\vspace{-0.5em}
\begin{quote}
\textbf{Example:} \\
\textbf{Input:} \textit{Click this link to provide delivery details.} \\
\textbf{Chain of Thought:} Suggests urgency and requests personal info --common phishing traits. \\
\textbf{Classification:} Scam
\end{quote}
\vspace{-0.5em}
By making reasoning explicit, CoT improves interpretability and error tracing.

\subsection{LLMs in Computer Security Tasks}

LLMs are increasingly used in computer security tasks. Fang \textit{et
al.}~\cite{fang2024} show they help in code analysis but struggle with
obfuscated logic. Liu \textit{et al.}~\cite{liu2024} find ChatGPT can outperform
some traditional tools in bug detection and repair, with prompt design being
key. Meng \textit{et al.}~\cite{meng2024} introduce \textsc{ChatAFL}, an
LLM-guided fuzzer that improves coverage and finds novel protocol bugs. Kande
\textit{et al.}~\cite{kande2023} use LLMs to generate security policies from
natural language.

For penetration testing, Deng \textit{et al.}~\cite{deng2024} propose
\textsc{PentestGPT}, a modular system that doubles task success rates via
chained prompting. Sufi \textit{et al.}~\cite{sufi2024} apply LLMs to threat
report summarization and prediction. Hu \textit{et al.}~\cite{hu2024llm} build
LLM-driven threat graphs for early attack detection. Together, these works show
LLMs’ growing role in automated security analysis.

\subsection{Scam/Fraud Detection and LLMs}

Recent studies cover a wide spectrum of scams: technical
support~\cite{miramirkhani2016dial,srinivasan2018exposing},
shopping~\cite{bitaab2023beyond,antiscam,carpineto2017learning,wabeke2020counterfighting,kharraz2018surveylance},
romance~\cite{suarez2019automatically,al2020social},
pet~\cite{price2020resource},
crypto/NFT~\cite{badawi2020automatic,das2021understanding},
tax~\cite{bidgoli2017hello}, survey~\cite{kharraz2018surveylance},
gaming~\cite{badawi2019game},
SMS~\cite{salman2024investigating,nahapetyan2024sms}, and social
media~\cite{chen2014scam,labonne2023spam}. Traditional ML approaches include
SVM, RF, NB, KNN, MLP, logistic regression, AdaBoost, and BERT-based models.
Deep learning models such as CNNs, LSTMs, and transformer-based classifiers also
show strong performance.

LLMs have recently been applied to phishing detection in websites and emails.
ChatPhishDetector~\cite{koide2024chatphishdetector},
PhishOracle~\cite{kulkarni2024ml}, and Knowphish~\cite{299740} use GPT-4V to
detect phishing via textual and visual cues. In email classification, GPT-3.5,
GPT-4, and fine-tuned models like CyberGPT~\cite{chataut2024can} outperform
older baselines. Rose \textit{et al.}~\cite{de2024hey} and Roy \textit{et
al.}~\cite{roy2024chatbots} show LLMs can also generate phishing content,
emulating brands and evasion tactics.

While prior work explores LLMs for phishing detection, it often focuses on
limited models and basic prompting (e.g., zero- or few-shot). In contrast, our
study evaluates a broader set of state-of-the-art LLMs and prompting strategies,
demonstrating and measuring the effectiveness of a fine-tuned model tailored for
scam detection.

Traditional ML models often assume training and test distributions
match~\cite{sommer2010outside}, limiting
generalizability~\cite{Lee2023MakingLL}. Contextual models such as BERT
generally outperform them in textual
tasks~\cite{GonzlezCarvajal2020ComparingBA,Gani2022FeatureEV,Guo2020IncorporatingBI}.
We include comparative benchmarks of LLMs vs. BERT to assess performance across
distributional shifts.

\subsection{Online Scam Definition and Detection}
\label{sec:problem}

An online scam is a form of social engineering that manipulates victims using
psychological tactics -- exploiting trust, fear, or urgency -- to steal money or
sensitive information. Scammers use channels like email, messaging apps, social
media, and phone calls to initiate contact, then lure victims to fake websites
or services. Common types include phishing, tech support fraud, investment
schemes, and romance scams. Scams typically unfold in two stages: initial
contact followed by exploitation, with some advanced scams involving prolonged
trust-building.

In this work, we define scam detection as the task of classifying a given input
into one of the specific categories (i.e., Scam, Clean, and Uncertain).

\section{Scam Data Collection}
\label{sec:scam_data_collection}

Collecting real-world scam data poses two main challenges: privacy concerns -- due
to PII exposure and victim embarrassment -- and the cross-platform nature of scams,
which limits centralized data access. To address this, we curated a multi-source
dataset combining public and proprietary data.

Each data point was labeled as \textit{Scam}, \textit{Clean}, or
\textit{Uncertain}. While the first two categories are standard, we introduced
the \textit{Uncertain} label to capture ambiguous cases where even expert
annotators cannot decide without external metadata (e.g., sender email, URL
context).

We further categorized these data points into 25 categories, emphasizing fine-grained classification to maximize the level of detail and specificity in the data.

\vspace{1mm} \textbf{Reddit.} We collected 4377 posts from the \textit{Is this a
scam?} flair of the \textit{Scams} subreddit~\cite{reddit} between January and
April 2024. These user-submitted posts typically contain textual descriptions
and attached screenshots. We applied OCR to images and integrated the extracted
text. After filtering incomplete entries, we retained 1270 data points: 716
labeled as \textit{Scam}, 118 as \textit{Clean}, and 436 as \textit{Uncertain}.

\vspace{1mm} \textbf{Google Images.} To supplement scam examples, we used
queries such as ``hi mom scams'', ``romance scam messages'', and ``fake delivery
SMS'' to gather 615 images. OCR was applied, and entries were manually labeled:
300 \textit{Scam}, 151 \textit{Clean}, and 164 \textit{Uncertain}.

\vspace{1mm} \textbf{MTurk.} We recruited participants on Mechanical Turk to
submit screenshots of legitimate messages (excluding PII). Of 857 submissions,
279 were labeled as \textit{Scam}, 376 as \textit{Clean}, and 202 as
\textit{Uncertain}.

\vspace{1mm} \textbf{Proprietary Dataset.} To validate our findings at scale, a
security company replicated our experiments on a dataset of 59,991 real-world samples. Labels include 31,081
\textit{Scam}, 9374 \textit{Clean}, and 19,536 \textit{Uncertain}.

Table~\ref{tab:scam_data} summarizes the number of labeled entries across all
sources. 

\subsection{Ethics}
\label{sec:ethics}

Our data collection focused exclusively on publicly available content from
Reddit, adhering strictly to Reddit’s Developer Platform Guidelines and the
updated Public Content Policy (effective May 9, 2024). We limited our scope to
scam-related discussions and excluded all personally identifiable information
(PII), such as usernames or email addresses, in compliance with platform
policies and ethical research standards.

We also applied Optical Character Recognition (OCR) to publicly available Google
Images to extract scam-related text. Only textual features relevant to scam
classification were retained; no PII was captured or stored.

All collected data was anonymized before human labeling. Annotators received
only the minimal content required for classification, without any extraneous
personal details. We implemented safeguards to preserve data confidentiality and
user anonymity throughout the process.

Our research followed a transparent, privacy-preserving methodology designed to
uphold ethical integrity while meeting scientific goals.

\begin{table}[t]
\footnotesize
\centering
\caption{Scam data.} 
\label{tab:scam_data}
    \begin{tabular}{lcccc}
    \toprule
        Data Source & Total Size & Scam \# & Clean \# & Uncertain \#\\
        \midrule
        Google images & 615 & 300 & 151 & 164 \\
        Reddit (Scam) & 1270 & 716 & 118 & 436 \\
        MTurk & 857 & 279 & 376 & 202 \\
        \midrule
        Total & 2742 & 1295 & 645 & 802 \\
        \toprule 
        Proprietary data & 59991 & 31081 & 9374 & 19536 \\
        \bottomrule
    \end{tabular}
\end{table}

\section{LLM Experiments in Scam Detection}
\label{sec:llm_methodology}


In the context of scam detection, LLMs can be queried directly by end users or
integrated into large-scale filtering systems. This is especially valuable when
users lack expertise in prompt design. In both settings, the model is expected
to classify the input into one of the predefined categories. Accordingly, we
define scam detection as the task of assigning a given input to one of three
classes: \Scam, \Clean, or \Uncertain.

Our evaluation considers the impact of model size, architecture, prompting
strategy, parameter tuning, and fine-tuning. Additionally, we train a BERT-based
classifier to compare the accuracy, generalization, and cost trade-offs between
LLMs and traditional models.

\subsection{Prompting}

Our primary goal is to classify inputs into one of three categories defined in
Section~\ref{sec:scam_data_collection}. To this end, we design prompts with two
key objectives: (1) consistency across different LLMs, and (2) clearly
structured outputs suitable for automatic evaluation.

Manual review is impractical at scale, so we enforce output formatting to enable
automated parsing. After extensive prompt engineering and iteration, we use a
final set of refined prompts for evaluation.

The simplest strategy is to directly instruct the LLM to classify the input as
\Scam, \Clean, or \Uncertain using a fixed output format. Below, we present the
main prompt used in our evaluations:

\begin{tcolorbox}[breakable, left=0mm, right=0mm, top=0mm, bottom=0mm, sharp corners]
\textbf{{\underline{1. Main Prompt}}} \vspace{0.1cm}

\footnotesize
Classify the following input into Scam, Clean, or Uncertain. Only print the classification result in the output format. \\
Output format: \\
\textless classification\_result\textgreater. \\
Input: 
\end{tcolorbox}

The main prompt uses a zero-shot approach with minimal instruction. Alternatively, we design a more detailed prompt that encourages reasoning by framing the task as assisting the user in determining whether a given text is a scam. The model is instructed to consider arguments for both scam and clean interpretations before making a classification. Below, we present the reasoning prompt:


\begin{tcolorbox}[breakable, left=0mm, right=0mm, top=0mm, bottom=0mm, sharp corners]
\textbf{{\underline{2. Reasoning Prompt - Chain of Thought}}} \vspace{0.1cm}

\footnotesize

You are an assistant that helps to determine whether a text is fraudulent or benign. Given the text below, answer honestly and correctly whether the text is a scam or not. The answer should be only one of the following words: SCAM, CLEAN, or UNCERTAIN, and should not contain anything else. Consider both arguments for and against the text being classified as a scam. First,  write the best argument that it is scam in \textless scam\_argument\textgreater\space tags, and the best argument it is clean in \textless clean\_argument\textgreater\space tags. Then, if you believe that the text is a scam or fraud, write \textless classification\textgreater SCAM\textless/classification\textgreater\space. If you believe that the text is clean or benign, respond with \textless classification\textgreater CLEAN\textless /classification\textgreater\space. Otherwise, if the two arguments are roughly equally convincing, write \textless classification\textgreater UNCERTAIN\textless /classification\textgreater. \\
Input: 
\end{tcolorbox}

The reasoning prompt follows a chain-of-thought strategy, where the LLM generates two intermediate arguments—one for classifying the input as scam, the other as clean—before producing the final classification. This structured reasoning provides a stronger foundation for the final decision.

\noindent \textbf{Auxiliary Content.} To enhance performance, we also provide external context that guides the model's attention. Even if the information is widely known, explicitly presenting it improves the model’s ability to reason.

For example, a list of common scam indicators helps the model focus on relevant patterns. By injecting this knowledge, we improve the likelihood of accurate classification. The corresponding prompt is shown below:

\begin{tcolorbox}[breakable, left=0mm, right=0mm, top=0mm, bottom=0mm, sharp corners]
\textbf{{\underline{3. Common Signs Prompt}}} \vspace{0.1cm}

\footnotesize

The common signs of a scam:
1. Unsolicited Contact
2. Too Good to Be True Offers
3. Urgency and Pressure
4. Requests for Sensitive Personal Information
5. Requests for unusual forms of payment
6. Official-Looking Communications
7. Requests for Payment to Receive Prizes
8. Untraceable Contacts
9. Poor Grammar or Spelling
10. Romantic Interest
11. High-Pressure Sales Tactics
12. Unverified Sources.
You are an assistant that helps to determine whether a text is fraudulent or benign. Using the common signs of a scam provided above, answer honestly and correctly whether the text given below is a scam or not. The answer should be only one of the following words: SCAM, CLEAN, or UNCERTAIN, and should not contain anything else. \\
Input: 
\end{tcolorbox}

URLs are a key feature in scam detection, as fraudsters often embed malicious links to direct users to phishing pages, impersonate legitimate websites, or distribute malware. Since LLMs lack built-in URL intelligence, we enhance classification accuracy by injecting external safety assessments into the prompt.

For each input, we extract and expand shortened URLs. We then filter out non-malicious domains using the Tranco top 10K list~\cite{pochat2018tranco}, treating them as likely safe. Remaining URLs are evaluated using ScamAdviser~\cite{scamadviser}, whose numeric trust scores are translated into textual verdicts based on their published rating scale\footnote{https://www.scamadviser.com/articles/scamadviser-algorithm-explainer}.

We integrate this intelligence directly into the prompt to inform the model’s decision. The prompt used is shown below:

\begin{tcolorbox}[breakable, left=0mm, right=0mm, top=0mm, bottom=0mm, sharp corners]
\textbf{{\underline{4. URL Intelligence Prompt}}} \vspace{0.1cm} \\
\footnotesize
    Classify the following input into Scam, Clean, or Uncertain. Only print the classification result in the output format. \\
    Output format: \\
    \textless classification\_result\textgreater. \\
    The given text might include a URL analysis. If the analysis indicates that one of the URLs is unsafe, classify it as SCAM.
    Input: \textless input\textgreater\\
    
    URL analysis results: \\
    URL: A \\
    Verdict: Very Likely Safe\\
    URL: B \\
    Verdict: Very Likely Unsafe
\end{tcolorbox}

\noindent \textbf{Few-Shot.} All previous prompts were zero-shot, providing no
examples. To evaluate the impact of in-context learning, we also construct
few-shot prompts by prepending $K$ labeled examples to the main prompt.

Example selection follows a controlled procedure: (1) we randomly sample $K$
inputs from the dataset and record them, (2) the same $K$ examples are reused
across models for consistency, and (3) each few-shot setup is repeated 10 times
to ensure statistical robustness. For instance, in the five-shot setting, five
examples are sampled and fixed across all models and runs. Below, we show the
few-shot prompt used:

\begin{tcolorbox}[breakable, left=0mm, right=0mm, top=0mm, bottom=0mm, sharp corners]
\textbf{{\underline{5. Few-Shot Prompt}}} \vspace{0.1cm} \\
\footnotesize
Classify the following input into Scam, Clean, or Uncertain. Only print the classification result in the output format. \\
Output format: \\
\textless classification\_result\textgreater. \\
Examples: \\
Input: \textless example\_input\textgreater \\
Output: \textless example\_classification\textgreater \\ \\
Now, classify the following input:\\
Input: \textless input\textgreater
\end{tcolorbox}

\subsection{Parameter Tuning}

Another important factor in classification performance is the level of
determinism in the LLM's output, which is influenced by its sampling
configuration. Determinism is inversely related to creativity: while low
randomness (i.e., high determinism) typically favors classification tasks by
promoting consistent, high-probability outputs, increased randomness may help in
more ambiguous cases—such as scam detection—by allowing exploration of less
likely but plausible interpretations.

Two key hyperparameters govern output stochasticity: \textit{temperature} and
\textit{top-p}. Temperature controls the sharpness of the probability
distribution over tokens; lower values make the model more deterministic, while
higher values encourage diverse, creative outputs. Top-p (nucleus sampling)
defines a cutoff for the cumulative token probability, limiting sampling to the
most likely subset of tokens. Larger top-p values introduce greater variability.

To assess the impact of randomness, we tune these parameters during evaluation.
For top-p, we select values $\{0.1, 0.5, 0.9\}$ to span the [0, 1] range. For
temperature, we use $\{0.1, 0.5, 0.9\}$ within the typical range [0, 2],
avoiding overly creative outputs that may reduce classification reliability.

\begin{table}[!t]
\footnotesize
\centering
\caption{Large Language Models in scope.}\label{tab:llm_models}
\begin{tabular}{lclc}
\toprule
\multicolumn{2}{c}{Models with parameter sizes} &
\multicolumn{2}{c}{Models with unknown parameter sizes} \\
\cmidrule(lr){1-2} \cmidrule(lr){3-4}
LLM & Params & LLM & Params \\
\midrule
Mistral & 7B, 8x7B & ChatGPT 4 Turbo & N/A \\
Llama 3 & 8B & ChatGPT 4o & N/A \\
Llama 3.1 & 8B, 70B & ChatGPT 4o1 & N/A \\
Llama 3.2 & 11B & & \\
\bottomrule
\end{tabular}
\end{table}

\subsection{Fine-Tuning}
\label{sec:fine-tuning}

Fine-tuning is a supervised process that adapts a pretrained model to a specific
task by adjusting its weights using new labeled data. This approach avoids the
computational cost of training from scratch while allowing the model to
specialize, often improving task-specific performance.

In our evaluation, we fine-tune smaller LLMs and assess their performance using
the main prompt. To construct a balanced training set, we randomly sample 100
and 400 examples per class. For validation and testing, we select 100 examples
per class for each set, resulting in 300 samples each.

We tune key training parameters for optimal performance. The number of epochs
ranges from 1 to 4, batch size from 16 to 32, and learning rate from
$1\text{e}{-5}$ to $5\text{e}{-5}$.

\subsection{BERT in Scam Detection}

BERT, an early transformer-based model, serves as a compact and cost-efficient
alternative to modern LLMs. While it lacks generative capabilities, it performs
well on tasks involving text understanding and classification when fine-tuned
for the target domain.

Given its efficiency and strong performance, BERT is a viable candidate for scam
detection. To evaluate this, we fine-tune BERT on the same dataset used for LLM
fine-tuning and compare its classification performance against that of the LLMs.

\subsection{Evaluation Metrics}

For evaluation, we use standard multi-class classification metrics. Given the
three-class setup, we report both \textit{micro}- and \textit{macro}-averaged
precision and recall.  Micro-averaging treats each instance equally, making it
suitable in cases of class imbalance. Macro-averaging, in contrast, gives equal
weight to each class by computing metrics per class and averaging them. This
provides a clearer view of performance across all categories.  The formal
definitions of these metrics are provided in Equations~\ref{eq:1} to~\ref{eq:5}.

\begin{figure}[h]
\small
\centering

\begin{equation}
\text{Precision}_{\text{micro}} =
\frac{\sum_{i=1}^{N} \text{TP}_i}
{\sum_{i=1}^{N} (\text{TP}_i + \text{FP}_i)}
\label{eq:1}
\end{equation}

\begin{equation}
\text{Recall}_{\text{micro}} =
\frac{\sum_{i=1}^{N} \text{TP}_i}
{\sum_{i=1}^{N} (\text{TP}_i + \text{FN}_i)}
\label{eq:2}
\end{equation}

\begin{equation}
\text{Precision}_{\text{macro}} =
\frac{1}{N} \sum_{i=1}^{N}
\frac{\text{TP}_i}{\text{TP}_i + \text{FP}_i}
\label{eq:macro_precision}
\end{equation}

\begin{equation}
\text{Recall}_{\text{macro}} =
\frac{1}{N} \sum_{i=1}^{N}
\frac{\text{TP}_i}{\text{TP}_i + \text{FN}_i}
\label{eq:macro_recall}
\end{equation}

\begin{equation}
\text{F1} =
\frac{2 \cdot \text{Precision} \cdot \text{Recall}}
{\text{Precision} + \text{Recall}}
\label{eq:5}
\end{equation}

\end{figure}

While our primary metrics cover all three classes, we also report results for
the \Scam{} and \Clean{} classes only, as these may be more relevant in
practical scenarios, for instance, when prioritizing scam detection in
real-world systems.

\section{Evaluation Results}
\label{sec:evaluation}

We now evaluate the LLMs as described in our methodology. Our evaluation includes a diverse set of both open-source and proprietary LLMs. Table \ref{tab:llm_models} presents the LLMs in scope of this work. The models are from three different vendors, including open-source models with parameter sizes ranging from 7B to 70B, and proprietary models with unknown parameter sizes. Although the exact parameter sizes for the proprietary models are not disclosed, it is known that their predecessor (i.e., GPT-3) has 175B parameters~\cite{brown2020language}; hence, they are likely to be larger.

To run these models, we use two NVIDIA A100 80GBs for open-source models and Generative AI APIs for proprietary models. We repeat each experiment five times unless otherwise stated to create reliable result. Then, we get the majority voting of them.

\begin{table*}[!ht]
\centering
\caption{Zero-Shot Prompts Evaluation Results. R: Recall, P: Precision, S: Scam, C: Clean, U: Uncertain. For example, $P_S$ represents the precision for scam category. }\label{tab:zero_shot_result}
\resizebox{\textwidth}{!}{
    \begin{tabular}{lcccccccccccccc}
    \toprule
        LLM model & Prompt Type & $P/R/F1_{micro}$ & $F1_{macro}$ & $P_{macro}$ & $R_{macro}$ & $F1_{S}$ & $P_S$ & $R_S$ & $F1_{C}$ & $P_C$ & $R_C$ & $F1_{U}$ & $P_U$ & $R_U$ \\
        \midrule
        Mistral 7B & Main Prompt & 0.470  &  0.452  &  0.485  &  0.423  &  0.610  &  0.639  &  0.585  &  0.185  &  0.471  &  0.115  &  0.429  &  0.344  &  0.570 \\
        Mistral 7B & Chain of Thought & 0.354  &  0.493  &  0.674  &  0.389  &  0.185  &  0.882  &  0.103  &  0.165  &  0.831  &  0.091  &  0.469  &  0.309  &  0.971 \\
        Mistral 7B & Common Signs & 0.518  &  0.515  &  0.552  &  0.482  &  0.644  &  0.696  &  0.599  &  0.314  &  0.590  &  0.214  &  0.468  &  0.371  &  0.632  \\
        Mistral 7B & URL Intelligence & 0.514  &  0.486  &  0.489  &  0.483  &  0.631  &  0.612  &  0.652  &  0.478  &  0.483  &  0.474  &  0.346  &  0.372  &  0.323 \\
        \midrule
        Mistral 8x7B & Main Prompt & 0.507  &  0.485  &  0.509  &  0.463  &  0.659  &  0.693  &  0.628  &  0.289  &  0.485  &  0.206  &  0.428  &  0.348  &  0.555  \\
        Mistral 8x7B & Chain of Thought & 0.540  &  0.496  &  0.564  &  0.442  &  0.717  &  0.608  &  0.874  &  0.169  &  0.689  &  0.096  &  0.374  &  0.394  &  0.357  \\
        Mistral 8x7B & Common Signs & 0.542  &  0.539  &  0.560  &  0.519  &  0.655  &  0.710  &  0.608  &  0.487  &  0.595  &  0.412  &  0.442  &  0.375  &  0.537  \\
        Mistral 8x7B & URL Intelligence & 0.496  &  0.517  &  0.530  &  0.505  &  0.564  &  0.756  &  0.450  &  0.492  &  0.484  &  0.499  &  0.432  &  0.349  &  0.566  \\
        \midrule
        Llama 3 8B & Main Prompt & 0.482  &  0.416  &  0.515  &  0.349  &  0.648  &  0.483  &  0.985  &  0.036  &  0.632  &  0.019  &  0.079  &  0.432  &  0.044  \\
        Llama 3 8B & Chain of Thought & 0.530  &  0.467  &  0.521  &  0.422  &  0.698  &  0.563  &  0.918  &  0.185  &  0.619  &  0.109  &  0.295  &  0.381  &  0.241  \\
        Llama 3 8B & Common Signs & 0.560  &  0.510  &  0.506  &  0.514  &  0.719  &  0.643  &  0.815  &  0.525  &  0.438  &  0.654  &  0.126  &  0.437  &  0.074   \\
        Llama 3 8B & URL Intelligence & 0.541  &  0.487  &  0.492  &  0.483  &  0.687  &  0.609  &  0.787  &  0.457  &  0.495  &  0.423  &  0.291  &  0.371  &  0.239  \\
        \midrule
        Llama 3.1 8B & Main Prompt & 0.497  &  0.432  &  0.516  &  0.371  &  0.656  &  0.499  &  0.958  &  0.050  &  0.586  &  0.026  &  0.203  &  0.462  &  0.130   \\
        Llama 3.1 8B & Chain of Thought & 0.522  &  0.456  &  0.499  &  0.419  &  0.696  &  0.574  &  0.882  &  0.126  &  0.548  &  0.071  &  0.336  &  0.375  &  0.304 \\
        Llama 3.1 8B & Common Signs & 0.574  &  0.530  &  0.573  &  0.493  &  0.704  &  0.571  &  0.918  &  0.495  &  0.607  &  0.419  &  0.225  &  0.543  &  0.142 \\
        Llama 3.1 8B & URL Intelligence & 0.509  &  0.482  &  0.489  &  0.475  &  0.635  &  0.644  &  0.626  &  0.408  &  0.464  &  0.364  &  0.392  &  0.357  &  0.435 \\
        \midrule
        Llama 3.1 70B & Main Prompt & 0.610  &  0.570  &  0.610  &  0.535  &  0.736  &  0.614  &  0.919  &  0.538  &  0.728  &  0.426  &  0.339  &  0.489  &  0.259  \\
        Llama 3.1 70B & Chain of Thought & 0.632  &  0.595  &  0.601  &  0.590  &  0.764  &  0.718  &  0.815  &  0.591  &  0.627  &  0.560  &  0.423  &  0.457  &  0.394   \\
        Llama 3.1 70B & Common Signs & 0.631  &  0.587  &  0.587  &  0.587  &  0.761  &  0.677  &  0.869  &  0.638  &  0.585  &  0.702  &  0.274  &  0.500  &  0.188  \\
        Llama 3.1 70B & URL Intelligence & 0.619  &  0.609  &  0.610  &  0.608  &  0.714  &  0.753  &  0.679  &  0.635  &  0.637  &  0.633  &  0.474  &  0.440  &  0.512 \\
        \midrule
        Llama 3.2 11B & Main Prompt & 0.504  &  0.453  &  0.542  &  0.388  &  0.666  &  0.522  &  0.917  &  0.048  &  0.696  &  0.025  &  0.289  &  0.410  &  0.223   \\
        Llama 3.2 11B & Chain of Thought & 0.513  &  0.433  &  0.472  &  0.400  &  0.692  &  0.554  &  0.921  &  0.110  &  0.506  &  0.062  &  0.269  &  0.355  &  0.217  \\
        Llama 3.2 11B & Common Signs & 0.571  &  0.526  &  0.567  &  0.491  &  0.703  &  0.570  &  0.916  &  0.498  &  0.605  &  0.423  &  0.213  &  0.525  &  0.133  \\
        Llama 3.2 11B & URL Intelligence & 0.508  &  0.487  &  0.493  &  0.481  &  0.625  &  0.649  &  0.602  &  0.426  &  0.469  &  0.391  &  0.401  &  0.361  &  0.451 \\
        \midrule
        ChatGPT 4 Turbo & Main Prompt & 0.573  &  0.542  &  0.606  &  0.490  &  0.740  &  0.668  &  0.829  &  0.192  &  0.740  &  0.110  &  0.463  &  0.410  &  0.531  \\
        ChatGPT 4 Turbo & Chain of Thought & 0.576  &  0.608  &  0.700  &  0.536  &  0.716  &  0.811  &  0.641  &  0.295  &  0.891  &  0.177  &  0.531  &  0.399  &  0.792  \\
        ChatGPT 4 Turbo & Common Signs &  0.631  &  0.593  &  0.622  &  0.567  &  0.746  &  0.626  &  0.922  &  0.632  &  0.707  &  0.572  &  0.298  &  0.534  &  0.207 \\
        ChatGPT 4 Turbo & URL Intelligence & 0.359  &  0.458  &  0.497  &  0.424  &  0.186  &  0.810  &  0.105  &  0.456  &  0.383  &  0.563  &  0.400  &  0.298  &  0.605   \\
        \midrule
        ChatGPT 4o & Main Prompt & 0.616  &  0.588  &  0.641  &  0.543  &  0.752  &  0.658  &  0.878  &  0.468  &  0.812  &  0.329  &  0.437  &  0.451  &  0.423  \\
        ChatGPT 4o & Chain of Thought & 0.598  &  0.589  &  0.682  &  0.518  &  0.773  &  0.715  &  0.841  &  0.242  &  0.918  &  0.140  &  0.480  &  0.412  &  0.575 \\
        ChatGPT 4o & Common Signs & 0.608  &  0.578  &  0.634  &  0.530  &  0.763  &  0.670  &  0.886  &  0.418  &  0.809  &  0.282  &  0.423  &  0.424  &  0.421 \\
        ChatGPT 4o & URL Intelligence &  0.527  &  0.550  &  0.570  &  0.531  &  0.611  &  0.801  &  0.493  &  0.509  &  0.544  &  0.478  &  0.459  &  0.364  &  0.622  \\
        \midrule
        ChatGPT 4o1 & Main Prompt & 0.641  &  0.612  &  0.649  &  0.579  &  0.769  &  0.658  &  0.924  &  0.638  &  0.705  &  0.583  &  0.329  &  0.584  &  0.229  \\
        ChatGPT 4o1 & Chain of Thought & 0.623  &  0.583  &  0.611  &  0.558  &  0.763  &  0.649  &  0.924  &  0.608  &  0.647  &  0.574  &  0.263  &  0.536  &  0.175  \\
        ChatGPT 4o1 & Common Signs & 0.626  &  0.592  &  0.630  &  0.559  &  0.762  &  0.645  &  0.931  &  0.617  &  0.696  &  0.553  &  0.284  &  0.548  &  0.192   \\
        \bottomrule
    \end{tabular}
    }
\end{table*}


\subsection{Zero-Shot Prompting}

Table~\ref{tab:zero_shot_result} presents the evaluation results for zero-shot prompts, and we now discuss each of them.




\noindent \textbf{Main Prompt.} The best performing model is ChatGPT 4o1, followed by ChatGPT 4o and Llama 3.1 70B in terms of its micro and macro precision, recall, and F1. Following this pattern, models with larger parameter sizes performed better than those with smaller parameter sizes, with the exception of Mixtral 8x7B, which performed comparably to the smaller models.

In \Scam category, five models achieved over 90\% recall, while precision topped out at 69\% at for all models. When we account for precision and recall together, ChatGPT 4o1 was the best performing model.

In general most of the models had hard time to categorize the \Clean samples, where the recall for some of them is as low as 0.019, meaning they were only able to detect 2\% of the all Clean samples. Compared to other models, ChatGPT 4o1 was the outstanding in term of its performance in \Clean category, where it had 58\% recall. Overall, again ChatGPT 4o1 performed better than other according to its F1 score.

Similarly, in \Uncertain category, many models struggled to classify the samples as uncertain. F1 scores were low and ChatGPT 4 Turbo had the best F1 score for \Uncertain category.




\noindent \textbf{Chain of Thought.} While the prompt mostly improved performance, it led to degradation in some models. For instance, Mistral 7B, ChatGPT 4o, and ChatGPT 4o1 performed worse compared to the main prompt in micro metrics. Interestingly, with this prompt, LLama 3.1 70B was able to surpass the performance of ChatGPT 4o, but its performance still fell short of ChatGPT 4o1. In terms of F1 macro, ChatGPT-4 Turbo's performance was generally the best.

The prompt significantly degraded Mistral 7B's F1 score for \Scam. For the others, there was an increase in the score, except for small decreases in ChatGPT-4 Turbo and ChatGPT-4o1.

For \Clean and \Uncertain F1 scores, there was a mixed response, with some models showing improved performance and others showing degraded performance.

Overall, when dealing with smaller models, this prompt can be particularly useful for improving performance, whereas in larger models, the effect on performance is mixed.



\noindent \textbf{Common Signs.} Except for ChatGPT 4o and ChatGPT 4o1, the prompt significantly improved the micro and macro metrics compared to the main prompt. For example, for the Llama 3.2 11B model, the main micro F1 value was 50\%, but it increased to 57\% with this prompt. The prompt also improved the F1 scores for \Clean significantly for some models. For instance, for Llama 3.1 8B, it increased from 5\% to 49\%. For other categories, there were mixed effects.

Llama 3.1 70B and ChatGPT 4 Turbo were the top performers with the prompt, achieving a 63\% micro F1 score. However, they were not able to surpass ChatGPT 4o1's performance with the main prompt. Overall, including the common signs in the prompt is an important factor, but it does not work the same way across all models.

\noindent \textbf{URL Intelligence.} Results for URL intelligence were not promising. There were cases where it improved performance for some models; however, there were always other prompts that worked better with the same model. The best-performing model was Llama 3.1 70B with a 0.619 micro F1 score.

Overall, the effect of prompts varies across LLMs. While some prompts work better for one model, they might not perform as well with others. Additionally, prompts can affect recall and precision metrics differently. In general, prompts should be tested with the specific model that will be used. Smaller models tend to benefit more from prompting, but still do not reach the performance level of larger models.

The performance across specific categories also varies. The best-performing model in terms of micro and macro F1 scores, ChatGPT 4o1, provides the best performance for the \Scam and \Clean categories; however, ChatGPT 4 Turbo performs best for \Uncertain.

Among the larger models, Llama 3.1 70B could not outperform, but performed comparably to proprietary models. This is also important because it demonstrates the possibility of using a local model with a 70B parameter size in scam detection tasks, achieving performance comparable to proprietary models.

\subsection{Many-Shot Prompting}

\begin{figure*}[t]
\centering
\includegraphics[width=\linewidth]{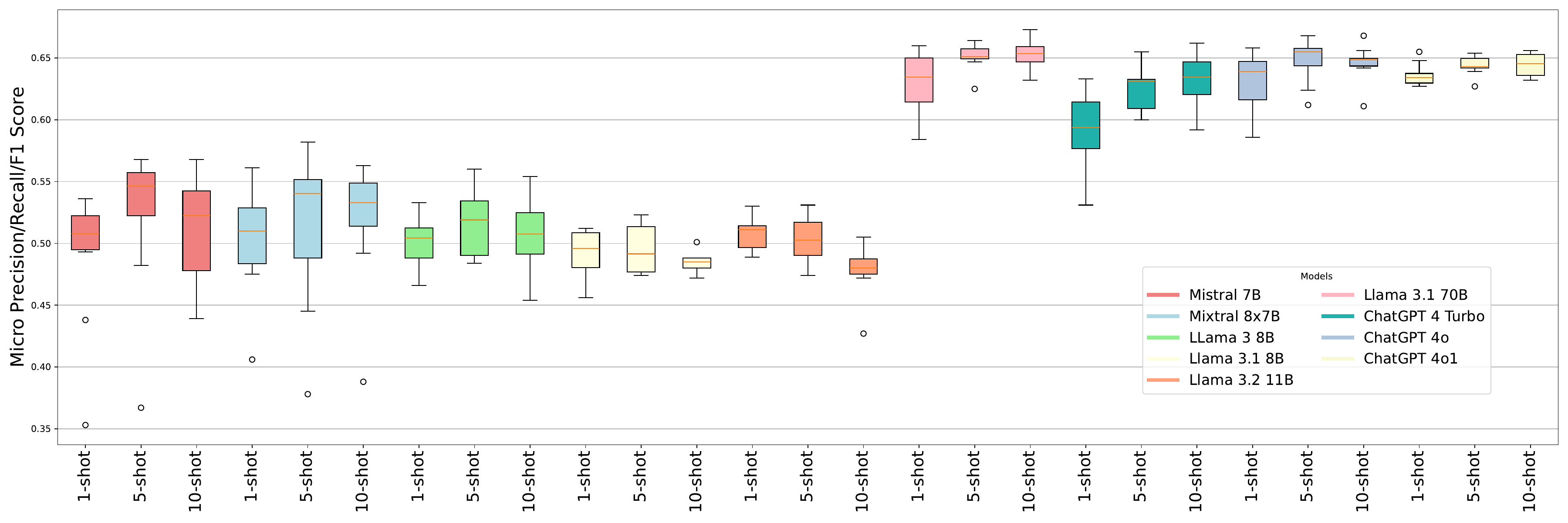}
\caption{Many-Shot Prompt Micro Precision, Recall, and F1 score across models, and 1, 5, and 10 shot.}
\label{fig:few_shot}
\end{figure*}

Figure~\ref{fig:few_shot} presents the micro precision, recall, and F1 scores across models and 1, 5, and 10-shot values.

This time, the best-performing model was Llama 3.1 70B, with a micro F1 of 0.653 for the 10-shot setting, surpassing the other models with different many-shot prompts and ChatGPT 4o1 with the main prompt. 

In general, having more shots increased the performance of the models. However, there were also cases where it negatively affected the model's performance. For example, in the case of Llama 3.2 11B, the best performance was with the one-shot prompt. Interestingly, some models performed better with 5 shots than with 1 or 10 shots. This is especially true for smaller models, which can handle shorter inputs more effectively.

Many-shot prompting is useful for improving model performance. However, for some models, zero-shot prompting can still yield better results. Additionally, the number of shots that performs best for each model can vary. Therefore, each model requires experimentation to determine which shot number will be the most effective.

\subsection{Temperature and Top-p}

As the next step, we evaluate models with temperature and top-p values of 0.1, 0.5, and 0.9. Since the main prompt has been shown to be a simple and effective strategy, we use it in this and the remaining sections of our paper for our experiments with LLMs. Table~\ref{tab:temperature-top-p} presents the results of this experiment.

Although there are minor changes in the micro precision, recall, and F1 scores, we did not observe any drastic changes. Therefore, based on initial experiments, we only tested the local models for temperature and top-p. Overall, there is no optimal temperature and top-p for all models; they should be individually evaluated, and the benefit is limited.
We also tested larger temperature values, which further decreased the F1 scores; therefore, we decided not to continue.

\begin{table*}[t]
\scriptsize
\centering
\caption{Micro Precision/Recall/F1 for Temperature x Top-p combinations across six models using the main prompt.}
\label{tab:temperature-top-p}

\begin{minipage}[t]{0.3\linewidth}
\centering 
\textbf{Mistral 7B} \\
\vspace{0.5em}
\resizebox{\textwidth}{!}{

\begin{tabular}{c|ccc}

Temp.\textbackslash{}Top-p & 0.1 & 0.5 & 0.9 \\
\midrule
0.1 & 0.470 & 0.470 & 0.470 \\
0.5 & 0.470 & 0.470 & 0.471 \\
0.9 & 0.470 & 0.468 & 0.470 \\
\end{tabular}
}
\end{minipage}
\hfill
\begin{minipage}[t]{0.3\linewidth}
\centering 
\textbf{Llama 3.2 11B} \\
\vspace{0.5em}
\resizebox{\textwidth}{!}{

\begin{tabular}{c|ccc}
Temp.\textbackslash{}Top-p & 0.1 & 0.5 & 0.9 \\
\midrule
0.1 & 0.503 & 0.503 & 0.503 \\
0.5 & 0.503 & 0.504 & 0.511 \\
0.9 & 0.504 & 0.504 & 0.507 \\
\end{tabular}
}
\end{minipage}
\hfill
\begin{minipage}[t]{0.3\linewidth}
\centering 
\textbf{Llama 3.1 8B} \\
\vspace{0.5em}
\resizebox{\textwidth}{!}{

\begin{tabular}{c|ccc}
Temp.\textbackslash{}Top-p & 0.1 & 0.5 & 0.9 \\
\midrule
0.1 & 0.498 & 0.498 & 0.498 \\
0.5 & 0.498 & 0.497 & 0.502 \\
0.9 & 0.498 & 0.500 & 0.484 \\
\end{tabular}
}
\end{minipage}

\vspace{1.2em}

\begin{minipage}[t]{0.3\linewidth}
\centering 
\textbf{Llama 3 8B} \\
\vspace{0.5em}
\resizebox{\textwidth}{!}{

\begin{tabular}{c|ccc}
Temp.\textbackslash{}Top-p & 0.1 & 0.5 & 0.9 \\
\midrule
0.1 & 0.482 & 0.482 & 0.482 \\
0.5 & 0.482 & 0.482 & 0.483 \\
0.9 & 0.481 & 0.480 & 0.479 \\
\end{tabular}
}
\end{minipage}
\hfill
\begin{minipage}[t]{0.3\linewidth}
\centering 
\textbf{Mixtral 8x7B} \\
\vspace{0.5em}
\resizebox{\textwidth}{!}{

\begin{tabular}{c|ccc}
Temp.\textbackslash{}Top-p & 0.1 & 0.5 & 0.9 \\
\midrule
0.1 & 0.509 & 0.509 & 0.509 \\
0.5 & 0.509 & 0.507 & 0.509 \\
0.9 & 0.509 & 0.509 & 0.510 \\
\end{tabular}
}
\end{minipage}
\hfill
\begin{minipage}[t]{0.3\linewidth}
\centering 
\textbf{Llama 3.1 70B} \\
\vspace{0.5em}
\resizebox{\textwidth}{!}{

\begin{tabular}{c|ccc}
Temp.\textbackslash{}Top-p & 0.1 & 0.5 & 0.9 \\
\midrule
0.1 & 0.610 & 0.610 & 0.610 \\
0.5 & 0.610 & 0.610 & 0.609 \\
0.9 & 0.610 & 0.606 & 0.612 \\
\end{tabular}
}
\end{minipage}

\end{table*}

\subsection{Agreement and Ensemble Models}

We now evaluate the agreement across models and whether we can use an ensemble model to improve performance.

\begin{figure}[t]
\centering
\includegraphics[width=0.8\linewidth]{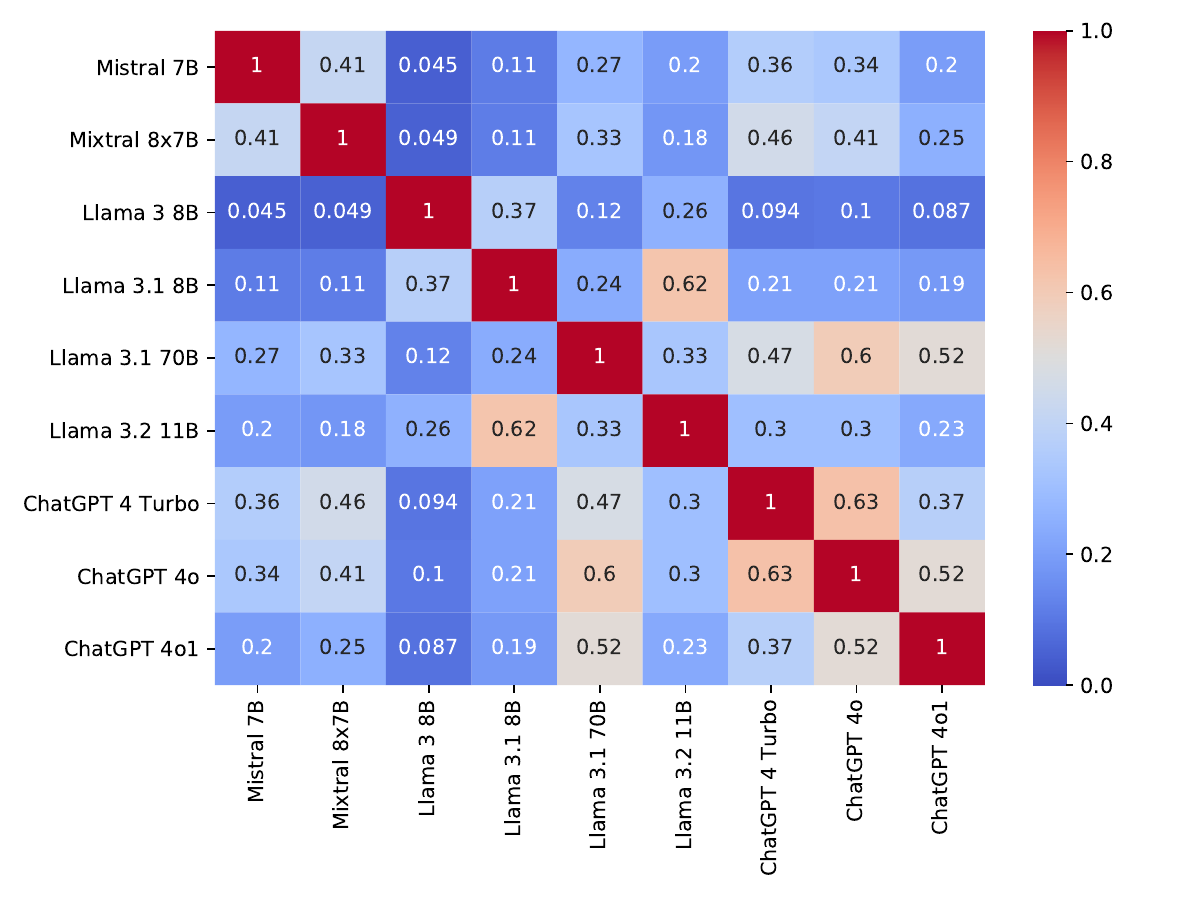}
\caption{Cohen's Kappa Scores, representing the agreement between models using the main prompt.}
\label{fig:kappa_scores}
\end{figure}

To calculate the agreement across models, we use Cohen's Kappa Score, a statistical measure used to assess inter-annotator agreement for categorical classification problems. A higher score means the models have more agreement, while a lower score means the opposite. Figure~\ref{fig:kappa_scores} presents the calculated scores across models.

In general, models that perform similarly also exhibit high agreement. For example, the ChatGPT models and Llama 3.1 70B show higher agreement within their own group but lower agreement with other models. Although these models have higher intra-group agreement, it is not perfect. This opens up the possibility of combining two or more models into an ensemble model.

To observe whether this is a potentially successful strategy, we test all different combinations of models. However, none were able to surpass the performance of the simple prompt.

\subsection{Fine-tuning}

\textbf{Fine-tuning LLMs.} We first fine-tune Mistral 7B and Llama 3.1 8B to explore whether fine-tuned smaller models can perform as well as larger models. As described in Section~\ref{sec:fine-tuning}, we used an equally and randomly distributed training, test, and validation set across three categories.

Table~\ref{tab:merged_finetuning_all} present the results for the fine-tuned models with various hyperparameters. The training set includes 400 samples per category -- 1200 in total. The test and validation sets contain 100 samples each, totaling 300.

For Mistral 7B, the fine-tuned model with 4 epochs, a batch size of 16, and a learning rate of 5e-5 performed the best, achieving a micro F1 score of 0.543. This is an improvement over the original 0.470 micro F1 score obtained with the main prompt. For Llama 3.1 8B, the same hyperparameters yielded the best performance, and similarly, fine-tuning helped improve the model's results.

While fine-tuning boosted model performance, it still fell short of larger models. Although gains are limited for smaller models, it could further improve performance in larger ones. Due to hardware and funding constraints, we were unable to attain these results.

\begin{table}[t!]
\centering
\caption{Fine-tuning results for Mistral 7B, Llama 3.1 8B, and BERT Micro F1 with varying hyperparameters using the main prompt.}
\label{tab:merged_finetuning_all}
\begin{tabular}{cccccc}
\toprule
\textbf{Epochs/Batch} & \textbf{Learning Rate} &
\textbf{Mistral 7B} & \textbf{Llama 3.1 8B} & \textbf{BERT} \\
\midrule
1/16 & 1e-5 & 0.443 & 0.490 & 0.460 \\
1/16 & 5e-5 & 0.357 & 0.430 & 0.540 \\
1/32 & 1e-5 & 0.423 & 0.313 & 0.360 \\
1/32 & 5e-5 & 0.350 & 0.460 & 0.480 \\
4/16 & 1e-5 & 0.510 & 0.413 & 0.510 \\
4/16 & 5e-5 & 0.543 & 0.553 & 0.590 \\
4/32 & 1e-5 & 0.463 & 0.480 & 0.510 \\
4/32 & 5e-5 & 0.510 & 0.520 & 0.580 \\
\bottomrule
\end{tabular}
\end{table}


\textbf{Fine-tuning BERT.} We now fine-tune a BERT base model and compare its performance with the LLMs we previously tested. We again use the same parameter ranges that were applied when fine-tuning the LLM models. The best performance was achieved using 4 epochs, a batch size of 16, and a learning rate of 5e-5, yielding a micro F1 score of 0.59.

The performance of BERT actually surpasses that of smaller LLMs and is only slightly lower than that of larger LLMs. Considering the costs associated with running LLMs, BERT appears to be a viable alternative.

Although BERT performed well on the test set after fine-tuning, its performance on larger datasets remains unclear. To evaluate this, we tested both the fine-tuned BERT model and Llama 3.1 70B using the company's proprietary data.

Table~\ref{tab:bert_real_data} presents the results on this proprietary dataset. While fine-tuned BERT performed better on the test set, it performed poorly here. In contrast, Llama 3.1 70B maintained a similar micro F1 score. These results indicate that, in terms of generalizability, LLMs offer significantly more. Therefore, they are particularly useful in the wild, where inputs are often unfamiliar or previously unseen.

\setlength{\tabcolsep}{1pt} 
\begin{table*}[!t]
\scriptsize
\centering
\caption{Data evaluation results for proprietary data. R: Recall, P: Precision, S: Scam, C: Clean, U: Uncertain. For example, P\_S represents the precision for scam category. }\label{tab:bert_real_data}
\resizebox{\textwidth}{!}{
    \begin{tabular}{lccccccccccccc}
    \toprule
        LLM model & $P/R/F1_{micro}$ & $F1_{macro}$ & $P_{macro}$ & $R_{macro}$ & $F1_{S}$ & $P_S$ & $R_S$ & $F1_{C}$ & $P_C$ & $R_C$ & $F1_{U}$ & $P_U$ & $R_U$ \\
        \midrule
        BERT fine-tuned & 0.393  &  0.412  &  0.402  &  0.422  &  0.557  &  0.629  &  0.500  &  0.319  &  0.208  &  0.683  &  0.137  &  0.368  &  0.084 \\

        
        Llama 3.1 70B & 0.596  &  0.525  &  0.523  &  0.527  &  0.757  &  0.675  &  0.860  &  0.453  &  0.417  &  0.497  &  0.305  &  0.476  &  0.225  \\

    \bottomrule
    \end{tabular}
    }
\end{table*}

\subsection{Scam Types}

Table~\ref{tab:misclassified_categories} presents the scam types that we annotated during our labeling process, and the number of correctly classified samples from each type by ChatGPT 4o1, which was proved to be the most effective for zero-shot prompting; thus, we further analyze its results.
Overall, ChatGPT 4o1 was able to correctly classify 90.35\% of the scams that are labeled as one of the scam types by the annotators. ChatGPT 4o1 was relatively less successful in Eshop and Phishing scams. This might be because these type of scams can be complex and include many URLs that are not immediately available.

\setlength{\tabcolsep}{3pt}
\begin{table*}[h!]
\scriptsize
\centering
\caption{Scam categories and the number of correct classifications by ChatGPT 4o1 using the main prompt.}
\label{tab:misclassified_categories}
\scriptsize

\resizebox{\textwidth}{!}{
\begin{tabular}{lrrr @{\hspace{1.8em}} lrrr}
\toprule
Scam Type & \# & Correct & \% & Scam Type & \# & Correct & \% \\
\midrule
Scareware scam               & 41  & 41  & 100.00 & Advance fee fraud       & 25  & 23  & 92.00 \\
Norton scam                  & 8   & 8   & 100.00 & Unexpected money        & 22  & 20  & 90.91 \\
Voice mail scam              & 8   & 8   & 100.00 & Investment scam         & 64  & 58  & 90.63 \\
Loan scam                    & 6   & 6   & 100.00 & Smishing scam           & 78  & 70  & 89.74 \\
Inheritance scam             & 6   & 6   & 100.00 & Lottery/Winning Scam    & 9   & 8   & 88.89 \\
Survey scam                  & 3   & 3   & 100.00 & Phishing scam           & 133 & 117 & 87.97 \\
Betting and sport investment & 2   & 2   & 100.00 & charity scam            & 7   & 6   & 85.71 \\
Insurance scam               & 1   & 1   & 100.00 & Impersonation scam      & 9   & 7   & 77.78 \\
Package delivery scam        & 121 & 120 & 99.17  & Goverment/Rebate scam   & 9   & 7   & 77.78 \\
Invoice scam                 & 84  & 83  & 98.81  & Romance/Sex scam        & 4   & 3   & 75.00 \\
Tech support scam            & 39  & 38  & 97.44  & Eshop Scam              & 92  & 67  & 72.83 \\
Jobs and employment          & 188 & 178 & 94.68  & Romance scam            & 15  & 14  & 93.33 \\
Lottery scam                 & 17  & 16  & 94.12  & Total                   & 1026& 927 & 90.35 \\
\bottomrule
\end{tabular}
}
\end{table*}

\section{Discussion}
\label{sec:discussion}

While generative AI and LLMs are widely adopted across domains, their superiority over classical machine learning (ML) models in structured tasks like classification remains debatable. In our study on scam detection, we found LLM performance to be highly sensitive to prompt design and model architecture. This suggests that despite their flexibility, LLMs struggle in domains where classical models are getting significant results.

A more pragmatic future lies in hybrid models that combine the semantic understanding of LLMs with the precision and efficiency of classical ML. For instance, LLMs can extract features or flag ambiguous inputs, while decision trees or ensemble models handle deterministic classification. Although BERT underperformed on out-of-domain data, retraining or domain-adaptive fine-tuning may close this gap without requiring massive-scale models. Future work should identify the parts of the classification pipeline where LLMs provide marginal gains and avoid deploying them as a complete solution. Optimal performance likely lies not in model size but in architectural synergy.

\section{Limitations}
\label{sec:limitations}

Scam data collection presents inherent challenges due to the presence of personally identifiable information (PII), which limits large-scale public sharing and aggregation. As a result, our dataset is confined to publicly available sources, primarily user-submitted examples on online forums and search engines. These are often posted by individuals seeking to verify suspicious messages. While our dataset is smaller and unbalanced than ideal, it is, to our knowledge, among the most curated and comprehensive collections dedicated to scam campaigns in the English language.

Model selection was constrained by both accessibility and cost. We included a diverse and representative set of LLMs based on popularity and architecture, covering multiple providers and sizes. However, larger proprietary models (e.g., GPT-o1-pro, Claude 3 Opus) were excluded from certain experiments due to high API costs. Additionally, each model was tested across five runs to account for response variability, but more runs would offer stronger statistical reliability. Results are also prompt-sensitive; while we used empirically effective prompts, different prompts may yield different outcomes -- a well-known limitation in LLM evaluations.

\section{Conclusion}
\label{sec:conclusion}
This study provides the first comprehensive benchmark evaluating the applicability of large language models (LLMs) for scam detection. We systematically analyzed multiple LLMs across various prompting methods, hyperparameters, and fine-tuning conditions, revealing insights into their performance, limitations, and deployment feasibility. Our results show that the larger variants of LLMs, such as ChatGPT 4o1 and LLama 3.1 70B, provide the best performance across LLMs; however, LLM's effectiveness is highly contingent on model-specific prompting and configuration. A more classical model such as BERT remains competitive in controlled settings but lacks the generalization capabilities of LLMs. Importantly, we demonstrate that no single model or approach universally solves scam detection. Instead, the path forward lies in hybrid architectures that combine the semantic strength of LLMs with the efficiency of classical ML methods. Our open-source dataset and fine-tuned models offer a foundation for reproducible research and future deployments of scam detection systems.

\section*{Acknowledgment}

This work was partly-supported by the National Science
Foundation under grant 2329540.

\bibliographystyle{plain}
\bibliography{paper}

\end{document}